\documentclass[11pt]{article}
\usepackage[utf8]{inputenc}
\usepackage{amsmath,amsfonts,amssymb,latexsym}
\usepackage[T1]{fontenc}
\newtheorem{satz}{Theorem}[section]

\newtheorem{assumption}[satz]{Assumption}

\newtheorem{bem}[satz]{Remark}

\newtheorem{conclusion}[satz]{Conclusion}
\newtheorem{ob}[satz]{Observation}

\newcommand{\mcal}{\mathcal}

\newcommand{\tit}{\textit}

\newcommand{\R}{\mathbb{R}}

\newcommand{\beq}{\begin{equation}}
\newcommand{\eeq}{\end{equation}}

\begin{document}
\thispagestyle{empty}
\begin{center}
\vspace*{1.0cm}
{\Large{\bf Observables need not be diffeomorphism invariant\\ in Classical and Quantum Gravity}}
\vskip 1.5cm

{\large{\bf Manfred Requardt}}

\vskip 0.5cm

Institut fuer Theoretische Physik\\
Universitaet Goettingen\\
Friedrich-Hund-Platz 1\\
37077 Goettingen \quad Germany\\
(E-mail: requardt@theorie.physik.uni-goettingen.de)

\end{center}

\begin{abstract}
The problem of observables in classical and quantum gravity is a long-standing one. It is sometimes argued that observable quantities should be diffeomorphsm invariant, following the philosophy of Dirac. We argue that diffeomorphism invariance in classical and quantum gravity is not primarily a case of gauge invariance but rather an example of spontaneous symmetry breaking of the diffeomorphism group. As a consequence, observables fall into two classes, Dirac observables which are invariant under the full diffeomorphism group and more general observables which take on different (expectation) values in the different phases of the broken (diffeomorphims) symmetry group. To this latter class belong for example scalar functions of the metric tensor.   

\end{abstract} \newpage
\setcounter{page}{1}
\section{Introduction}
The problem of observables in classical and quantum gravity has generated a lot of discussions in the more recent past. It is, for example explicitly addressed in \cite{Rov1} and the companion paper \cite{Rov2}. Historically the problem is closely related to the so-called \tit{Einstein hole problem}, i.e. to some of the consequences of general covariance in general relativity (GR). To mention a few representative references, see e.g. \cite{Sta1},\cite{Sta2},\cite{No1},\cite{No2},\cite{Mac},\cite{Gaul} and also \cite{Rov1}.

The central question is the physical meaning of the points of the event manifold underlying GR. In contrast to pure mathematics this is a non-trivial point in physics. While in pure diffeential geometry one simply decrees the existence of, for example, a (pseudo-) Riemannian manifold with a differentiable structure (i.e., an appropriate cover with coordinate patches) plus a (pseudo-) Riemannian metric, $g$, the relation to physics is not simply one-one. In textbooks about GR it is frequently stated that all diffeomorphic (space-time) manifolds, $\mcal{M}$ are physically indistinguishable (see e.g. \cite{Ellis}). Put differently:
\beq S-T=Riem/Diff    \eeq
(cf. e.g. the discussion in \cite{Requ1}, which will become relevant in the following).

This becomes particularly virulent in the \tit{Einstein hole problem}. I.e., assuming that we have a region of S-T, free of matter, we can apply a local diffeomorphism which only acts within this \tit{hole}, letting the exterior invariant. We get thus in general two different metric tensors
\beq g(x)\quad , \quad g'(x):=\Phi_{\ast}\circ g(x)   \eeq
in the hole while certain inital conditions lying outside of the hole are unchanged, thus yielding two different solutions of the Einstein field equations (cf. e.g. the discussion in \cite{Rov1}).  

Many physicists consider this to be a violation of determinism (which it is not!, se below) and hence argue that the class of observable quantities have to be drastically reduced  in (quantum) gravity theory. They follow the line of reasoning developed by Dirac in the context of \tit{gauge theory}, thus implying that GR is essentially also a gauge theory (see the discussion in \cite{Rov1} and \cite{Requ1}). This then winds up to the conclusion:
\begin{conclusion}(Dirac) observables in (quantum) gravity are quantities which are diffeomorphism invariant with the diffeomorphism group, $Diff$ acting from 
$\mcal{M}$ to $\mcal{M}$, i.e.
\beq \Phi:\mcal{M}\rightarrow  \mcal{M}   \eeq
\end{conclusion}
\begin{ob}One should note that with respect to physical observations there is no violation of determinism. As we already emphasized in \cite{Requ1}, an observer can never really observe two different metric fields on one and the same space-time manifold. This can only happen on the mathematical paper. He will use a fixed measurement protocol, using rods and clocks in e.g. a local inertial frame where special relativity locally applies and then extend the results to general coordinate frames. In the following (see also \cite{Requ1}) we discuss this aspect in the context of spontaneous symmetry breaking.
\end{ob} 

We get a certain orbit under $Diff$ if we start from a particular manifold $\mcal{M}$ with a metric tensor $g$ and take the orbit
\beq \{\mcal{M},\Phi_{\ast}\circ g\}   \eeq
In general we have additional fields and matter distributions on $\mcal{M}$ which are transformd accordingly.
\begin{ob}Note that not even scalars are invariant in general in the above sense, i.e., not even the Ricci scalar is observable in the Dirac sense:
\beq R(x)\neq \Phi_{\ast}\circ R(x)  \eeq
in the generic case. Thus, this would imply that the class of admissible observables can be pretty  small (even empty!). Furthermore, it follows that points of $\mcal{M}$ are not a priori distinguishable.
\end{ob}
On the other hand, many authors consider the Ricci scalar at a point to be an observable quantity. 

This discussion winds up to the question whether GR is a true gauge theory or perhaps only apparently so at a first glance, while on a more fundamental level it is something different. Note the discussion of Kuchar and DeWitt at the Osgood Hill Conference 1988, mentioned by Rovelli in \cite{Rov1}, see also \cite{Sta1}, loc.cit. the discussion after the talk by Rovelli:
\begin{quote}(Kuchar): Quantities non-invariant under the full diffeomorphism group are observable in gravity.
\end{quote}
The reason for these apparently diverging opinions stems from the role reference systems are assumed to play in GR with some authors arguing that the gauge property of general coordinate invariance  is only of a formal nature.

In the hole argument it is for example argued that it is important to add some particle trajectories which cross each other, thus generating concrete events on $\mcal{M}$. As these point events transform accordingly under a diffeomorphism, the distance between the corresponding coordinates $x,y$ equals the distance between the transformed points $\Phi(x),\Phi(y)$, thus being a Dirac observable. On the other hand, the coordinates $x$ or $y$ are not observable. 
\begin{bem}One should note that this observation is somewhat tautological in the realm of Riemannian geometry as the metric is an absolute quantity, put differently (and somewhat sloppily), $ds^2$ is invariant under passive and by the same token active coordinate transformation (diffeomorphisms) because ,while conceptually different, the transformation properties under the latter operations are defined as in the passive case.
\end{bem}
In the case of GR this absolute quantity enters via the \tit{equivalence principle}. I.e., distances are measured for example in a local inertial frame (LIF) where special relativity holds and are then generalized to arbitrary coordinate systems.

We think, the observations we made in \cite{Requ1} make it possible to give this question a new twist in favor of the point of view that gauge invariance is not! a relevant criterion.
\section{The Space-Time Manifold as a reconstructed derived Object}
We start our analysis with a brief discussion how a space-time manifold is actually (re)constructed as an idealized abstract object in GR. We think it is useful to emphasize the following point because in present day differential geometry just the opposite point of view is usually emphasized. 
\begin{ob}In (mathematical) differential geometry the coordinate-free method is strongly favored and has also influenced the modern representation in the physics textbooks. I.e., coordinates and coordinate methods are frequently considered to be of a somewhat inferior value.
\end{ob}

We think in physics another, dual, point of view should also be discussed, i.e., manifolds as abstracted from so-called \tit{identifcation spaces}. Let us start from a cover of coordinate patches, $\{U_i\}$, which result from the introduction of a class of overlapping reference systems and observers. We can then define \tit{identification maps}, $ \phi_j^i$, from a subset $U_j^i\subset U_j$ to a subset $U_i^j\subset U_i$, i.e. in our case, the overlap region  with $ \phi_j^i$ being homeomorphisms. We then take the \tit{disjoint sum} $\bigsqcup U_i$ and define an equivalence relation on  $\bigsqcup U_i$:
\beq U_i\ni x\sim y\in U_j   \eeq
if there exists a  $ \phi_j^i$ of the above kind with $\phi_j^i(y)=x$. 

This new space of equivalence classes we call $\mcal{M}$ and endow it with the canonocal quotient topology induced by the canonical map:
\beq \bigsqcup U_i\rightarrow\mcal{M}   \eeq
I.e., a set, $\mcal{O}\subset\mcal{M}$ is open if $\phi^{-1}(\mcal{O})$ is open in $\bigsqcup U_i$. This means that each $\phi^{-1}(\mcal{O})\cap U_i$ is open in $U_i$.
\begin{ob}We see from this construction, which is nearer to concrete laboratory physics, that $\mcal{M}$ is not the initial object in physics to start from but rather a reconstructed and derived abstract structure of a somewhat indirect ontological existence.
\end{ob} 
\section{The spontaneously broken Diffeomorphism Group  in (Quantum) Gravity and the Role of the Metric Field}
If GR is really a full-fledged gauge theory in the strict sense of the word, then the arguments in favor of Dirac observables are inescapable. However, in \cite{Requ1} we developed a different point of view. We compared the situation of diffeomorphism invariance in GR with the phenomenon of spontaneous symmetry breaking (SSB) in other fields (as e.g. many-body theory) and showed that they are completely equivalent. 

Briefly resuming the results derived in \cite{Requ1} we arrived at the following conclusion:
\begin{ob}The order parameter (field) in GR is the metric field, $g$. I.e., $Diff$ is spontaneously broken if a non-vanishing $g$ exists. That means, the symmetric phase is a space (or manifold) without a physically acting metric field. The different non-vanishing metric fields resulting from the application of $Diff$ on some $g$, describing our ordinary S-T, lable the different broken phases (in this particular language). The Goldstone excitations are the gravitons.
\end{ob}

In contrast to the picture of GR as a gauge theory and \tit{admissible observables} of the Dirac type, we now have the following:
\begin{ob} The fully symmetric phase (vanishing $g$) corresponds to the situation in which GR behaves as a gauge theory.
\end{ob}
In order to apply the methods of SSB appropriately to this field we have to regard classical GR as a coarse-grained \tit{effectice theory} of an underlying more fundamental quantum theory of S-T (called quantum gravity). While this latter theory does not yet exist in a fully developed form, we will need in the follwing only very few concepts which should occur in any future theory.
\begin{assumption}We assume that the presumed underlying quantum theory of S-T is a theory of many microscopic degrees of freedom (DoF). There exists a quantum observable, $\hat{g}(x)$, the expectectation value of which is the classical metric field, $g(x)$
\beq <\hat{g}(x)>:=g(x)   \eeq
\end{assumption}
\begin{bem} $\hat{g}(x)$ itself can possibly be an average over a certain cluster of more microscopic DoF.
\end{bem}

While in the symmetric, unbroken phase we have
\beq <\hat{g}(x)>=0   \eeq
in the broken phases (which represent our classical S-T of ordinary GR) it holds on the other hand:
\beq  <\hat{g}(x)>\neq 0  \eeq
with the rhs depending on the particular phase.

In \cite{Requ1} we discussed SSB in GR  in the local version, which is, furthermore, a little bit nearer to the situation in ordinary SSB in e.g. many-body physics. In the local version the gauge group, $Diff$, is replaced by $GL(n,\R)$, which represents the action of the functional matrix, $\partial y/\partial x$ of the diffeomorphism $\Phi$ on the various tensor quantities of the theory.  It acts in particular on the local frames of the corresponding frame bundle (cf. \cite{Requ1} for more details). We select local inertial frames (LIF) in which gravity locally vanishes and which transform into each other via the Lorentz group, $L_+^{\uparrow}(n,\R)$. 
\begin{ob}The coset space 
\beq GL(n,\R)/L_+^{\uparrow}(n,\R)   \eeq
labels the manifold of broken phases and is in a one-one correspondence with the metric field $g(x)$.
\end{ob}

One should note that the phenomenon of SSB in GR is more involved as e.g. in (quantum) many-body physics. There exist two options. We can employ the group $Diff$ and as the subgroup of unbroken or conserved symmetries with respect to a given order parameter field $g$ we take the isomorphisms relative to $g$, i.e., the transformations $\Psi$ with
\beq \Psi_{\ast}\circ g(x)=g(x)   \eeq
\begin{bem}Note that this subgroup can in general be very small, depending on the properties of the used metric.
\end{bem}
For the transformed metric fields, lying in the orbit of $g$, the invariance group is then made up of
\beq  \Phi\circ\Psi\circ\Phi^{-1}   \eeq
This is exactly as in ordinary SSB. I.e., the conserved symmetries leave the order parameter invariant and the different order parameters have corresponding invariance groups as in the above case. 

On the other hand, we can employ the local point of view in which we consider instead the local action of the symmetries in the bundle formalism.
In this approach slightly different and perhaps more fundamental properties are elucidated. One should note that the second point of view is not simply the restriction of the first approach to the local situation. The two viepoints are related but not completely identical. This holds in particular with respect to the technical details. 
\begin{ob}The crucial property in GR is the existence of LIF in the presence of a pseudo-Riemannian metric. In these LIF a local observer does not experience a gravitational field. This gravitational field is however present in the generic case in non-inertial reference frames.
\end{ob}
The following remark, while being seemingly obvious, is made to characterize again the difference between the unbroken, symmetric phase and the situation where we have a manifold of broken phases.
\begin{conclusion}We infer that the symmetric phase (vanishing $g$-field) is the phase where a gravitational field is generally absent while with a non-vanishing $g$-field it is only absent in the LIF with their restricted symmetry group of Lorentz transforms.
\end{conclusion} 

From our experience with SSB in the many-body regime we know that the emergence of a non-vanishing order parameter implies the corresponding emergence of long-lived collective Goldstone modes. In our case we identify the \tit{gravitons} as these \tit{collective excitations}. In these collective excitations many individual microscopic DoF act coherently together, that is:
\begin{conclusion}The existence of a classical non-vanishing metric field, $g(x)$, implies that the whole underlying \tit{quantum vacuum}, QX, changes its microscopic fine structure compared to the case with $<\hat{g}(x)>=0$, that is, the symmetric case.
\end{conclusion}

For our general discussion of admissible observables in (quantum) gravity theory we can now conclude the following. If diffeomorphism invariance in GR is in fact a case of SSB and not a pure gauge phenomenon, all observables, occurring in the symmetric phase, remain admissible observables in the broken phases while their expectation values become in general phase dependent. Furthermore, it is possible that in the broken phases more (classical) Dirac observables will exist as a consequence of the non-vanishing of the order parameter field $g$. Take for example the distance between events, discussed above. Without a metric, this concept does not make sense.
\begin{conclusion}In the broken phases, i.e., $<\hat{g}(x)>=g(x)\neq 0$, we have two classes of observables, a smaller class of Dirac observables which is still invariant under the full symmetry group $Diff$, and the class of admissible observables which are not invariant under the full diffeomorphsm group but for example under the smaller subgroup of isomorphisms of the particular metric tensor $g$ (the conserved symmetries in the language of SSB). All scalar functions of $g$ belong to this latter class. In the local picture the description is a little bit different, a particular role being played by the Lorentz group.
\end{conclusion}

These latter observables are fully admissible, in contrast to the opinion of quite a few experts in the field as can be seen by the following example taken from ordinary SSB. Take e.g. a spin system with the broken phases given by the orientation of the \tit{spontaneous magnetization}
\beq  \vec{m}(x):=<\vec{S}(x)> \quad\text{with}\quad  \vec{m}(x)=  const  \eeq
 With $\vec{e}(x)$ the unit vector in the $\vec{m}(x)$-direction the rotations about  $\vec{e}(x)=\vec{e}$  are unbroken and leave the expectation of all observables invariant, i.e.:
\beq   <R_{\vec{e}}(A)>=<A>    \eeq
while a general rotation rotates also the direction of $\vec{m}$:
\beq   <R(\vec{S})>\neq<\vec{S}>    \eeq
\begin{bem}Warning: the technical details are actually not that simple. In the broken symmetry case the broken symmetries or their generators are not well-defined operators in the Hilbert spaces under discussion (for more details see \cite{Requ2} and the references given there). Their existence is only a formal one.
\end{bem}
\section{Conclusion}
We argued that \tit{diffeomorphism invariance} should rather be considered as a case of SSB with the gravitons as Goldstone modes (cf. \cite{Requ1}). This is consistent with the lesson (taught e.g. by Einstein) that GR should be a generally covariant theory and that all
\beq (\mcal{M},g)\; ,\; (\Phi(\mcal{M}),\Phi_{\ast}(g))     \eeq
are physically indistinguishable. We explained this in detail in \cite{Requ1}. That is, for internal observers in the case of SSB (i.e., for observers living within  the various broken phases) everything looks identical provided they use adapted coordinate systems. That means, if the various observers use coordinate systems which are transformed accordingly. An external observer, using a fixed coordinate system, observes the different (expectation) values of spontaneously broken observables. In case of diffeomorphism invariance these are the various transformed metric tensors, taken at the same point in $\mcal{M}$. The former case applies if the points in $\mcal{M}$ are transformed as well. That is, there are observables in classical and quantum gravity which are not! diffeomorphism invariant.

\end{document}